# Parallel-Plate Waveguides for Terahertz-Driven MeV Electron Bunch Compression

Mohamed A. K. Othman[*], Matthias C. Hoffmann, Michael K. Kozina, X. J. Wang, Renkai K. Li, and Emilio A. Nanni

*SLAC National Accelerator Laboratory, Stanford University, Menlo Park, CA, 94025 USA.*
*[*mothman@slac.stanford.edu](mailto:mothman@slac.stanford.edu)*

**We demonstrate the electromagnetic performance of waveguides for femtosecond electron beam bunch manipulation and compression with strong-field terahertz (THz) pulses. The compressor structure is a dispersion-free exponentially-tapered parallel-plate waveguide (PPWG) that can focus single-cycle THz pulses along one dimension. We show test results of the tapered PPWG structure using electro-optic sampling (EOS) at the interaction region with peak fields of at least 300 kV/cm given 0.9 µJ of incoming THz energy. We also present a modified shorted design of the tapered PPWG for better beam manipulation and reduced magnetic field as an alternative to a dual-feed approach. As an example, we demonstrate that with 5 µJ of THz energy, the PPWG compresses a 2.5 MeV electron bunch by a compression factor of more than 4 achieving a bunch length of about 18 fs.**

## 1. Introduction

Probing ultrafast dynamics reveals unprecedented details in energy-matter interactions at the atomic scale of materials. Ultrafast electron diffraction (UED) [1–4], X-ray free-electron lasers (XFELs) [5–7], and ultrafast electron microscopy [8] are utilized to access such ultrafast processes. In ultrafast pump-probe experiments, the dynamics are initiated by a pump laser, then the sample is probed with an ultrashort laser pulse, electron bunch or x-ray pulse. The opportunities for ultrafast diffraction have widened due to the rapid development of sources, detectors, and instrumentation which allow for studying matter in extreme conditions [9].

Laser-generated broadband THz radiation is a powerful tool for probing, controlling, and interacting with matter [10]. In particular, THz pulses can manipulate the state of matter through resonant, and non-resonant interactions, without altering the electronic density of states [11]. Optoelectronic and all-optical techniques are commonly employed for generation and detection of pulsed THz radiation such as for example optical rectification in nonlinear crystals. Such technology enables broadband THz pulses, giving rise to extremely high electric fields [12–14]. THz pulses with a typical duration of about1 ps duration can further be utilized to generate, manipulate or accelerate electron beams [15–20]. In contrast, development of high power, narrowband THz sources may enable the next generation high efficiency MeV linear accelerators thanks to improved shunt impedance and small form factor [21–23].

High brightness beams are typically generated using high gradient radio frequency (rf) guns through photo emission. A promising approach for enhancing timing resolution in UED pump-probe experiments is to manipulate the electron beam using THz electromagnetic radiation intrinsically synchronized with pump lasers prior to sample interaction. Previously, experimental work was done to demonstrate attosecond electron diagnostics using laser-generated single-cycle THz pulses [24–26]. This technique offers beam-to-laser timing characterization with the potential to achieve sub-femtosecond accuracy, and at the same time a direct measurement of the bunch length with a sub-femtosecond resolution. However, time-resolution achieved in typical single shot MeV UED experiments is limited by the bunch length, and timing jitter between the pump laser and photoemission [2,25].

Therefore, beam compression is an essential step towards pushing the frontier of timing resolution in ultrafast experiments, and it is conventionally done with rf cavities with dispersive beamlines [27,28]. The main drawback is phase jitter in the rf cavity that is converted into timing jitter after compression limiting bunch lengths to several 10s of fs, besides the effective large size and cost. Some THz-driven techniques for beam compression were recently proposed [29–31]. One advantage of THz-driven compression is the reduction of timing jitter since the laser generated THz is which is also synchronized with photoemission. However, [29,31] were designed for keV beams, work on single or few electron bunches and lack optimized coupling of THz to the e-beam. Furthermore, [30] utilizes a high charge bunch for generating THz wakefields in which the latter cannot improve the timing jitter of the compressed electron beams relative to an external ultrafast laser. In this letter, we introduce a THz-driven compressor structure based on a dispersion-free tapered PPWG that is suitable for synchronous interaction with MeV bunches. To our knowledge, we demonstrate here the first optimized high field single-feed THz structure for beam compression, in contrast to the dual-feed scheme in the conventional PPWG structures [19]. In fact, this is the first study with direct measurement of high fields at the interaction region of a THz accelerating structure. In Sec. 2 we present the design of the PPWG and in Sec. 3 we show EO measurements of the THz fields in the structure, while in Sec. 4 we show a shorted variation that could potentially eliminate transverse dispersion cause by the THz propagation as well as improve the performance of the PPWG compressor.

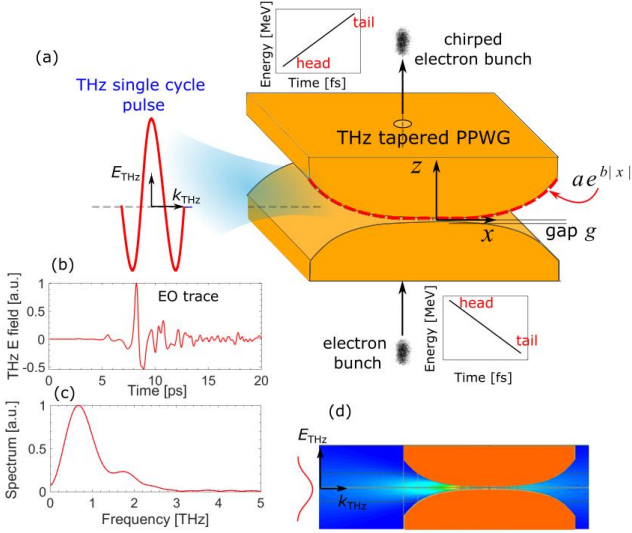

Fig. 1. (a) Terahertz tapered parallel-plate waveguide for MeV electron bunch compression. (b) Example of single cycle THz waveform measured using EOS in free space (with a 100 μm thick GaP crystal) and its spectrum in (c) that is coupled into the structure. (d) The resulting electric field profile of the coupled THz electric field $E_z$ field at a single frequency (0.65 THz).

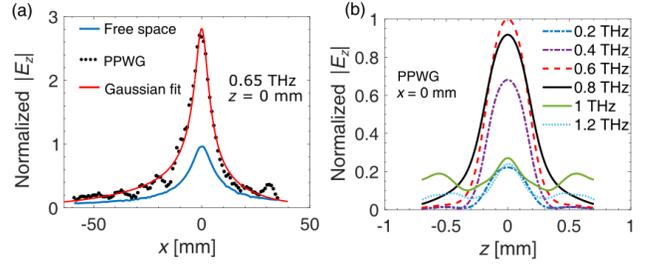

Fig. 2. Normalized electric field amplitude along the PPWG excited by a Gaussian beam assuming $g$ = 300 μm and $r_b$ = 125 μm. (a) Electric field amplitude along the $x$-direction versus that of the Gaussian beam in free space along the $x$ direction. (b) Electric field amplitude in the beam tunnel along the $z$-direction.

## 2. THz Tapered Parallel-Plate Waveguide

Parallel-plate waveguides are widely used in broadband THz-based applications [32–34], and there has been recent interest on utilizing PPWGs [17,19] for acceleration and beam manipulation. Here we investigate optimized, single-feed tapered PPWGs for MeV electron bunch compression. In Fig. 1(a) the tapered PPWG is shown and used to focus intense single-cycle THz pulses. The PPWG has a minimum gap $g$ with an exponentially-tapered surface viz. $a\exp(b|x|)$ where $x$ is the direction of the PPWG tapering as well as the wavevector of the incoming THz. An exponentially-tapered profile is chosen because it has been demonstrated to provide excellent coupling to broadband signals [35]. The PPWG parameters are set as $a = g/2$ and $b = 140$ m$^{-1}$ to provide optimal focusing to an incident broadband THz Gaussian beam. Note that the tapered PPWG only focuses in one dimension (along $x$) without disturbing the pulse propagation along this direction. Electron bunches enter the tapered PPWG compressor through a round beam tunnel and interact with the THz at the gap; providing sufficient energy chirp to compress the bunch after a certain drift distance. An illustration of the compression concept is also shown in Fig. 1 where an incoming electron beam bunch has a longitudinal phase space that has the head of the electron bunch traveling with higher velocity than its tail, causing the bunch to spread out. This bunch is manipulated by the PPWG which delivers an additional chirp, *i.e.*, flips the phase space, allowing the electrons in the tail to catch up, and eventually compress the bunch after some drift distance. In order to model the PPWG, full wave simulations of the structure were performed with the finite-element method implemented in *Ansys Electronics* for a frequency range from 0.05 THz to 1.5 THz, covering almost all the frequency spectrum of the THz pulse of interest shown in Fig. 1(c), with 0.65 THz being the central frequency of THz pulse. We assume a symmetric Gaussian beam with a minimum waist of 1 mm at $x = y = z = 0$ mm at 0.65 THz, which is found experimentally to provide the best coupling, i.e., highest field enhancement (See Sec. 3). Note that the minimum waist scales with frequency as $\propto 1/\sqrt{f}$ and this is considered in the broadband simulations. In Fig. 1(d) we show the field profile of the electric field in the tapered PPWG at 0.65 THz, which features no reflection from the tapered PPWG and demonstrates good coupling of the incoming Gaussian beam. We also report in Fig. 2(a) the field profile along the PPWG (along the $x$-direction) comparing both the incident Gaussian beam (analytical) and the total field focused by the PPWG (simulated) which is also fitted to a Gaussian profile. The electric field across the interaction gap at different frequencies along the beam axis ($z$-direction), is shown in Fig. 2(b). Here the field amplitudes are scaled by the spectrum of the pulse in Fig. 2(c) to provide the broadband single cycle pulse. The fields peak at the gap and decay away along the beam tunnel in the $z$-direction. Note that the cylindrical beam-tunnel cross section should be minimized to prohibit field leakage inside the tunnel and improve the field enhancement in the gap. We choose $r_b$ =125 μm as the beam tunnel diameter for practical implementation. Accordingly, the cutoff frequency of the beam tunnel for the fundamental mode is 0.92 THz allowing only the higher frequency content of the pulse to leak inside the tunnel and experience less field enhancement as shown from the fields in Fig. 2(b) above 1 THz.

## 3. THz Characterization Using EO Sampling

In this section we present the experimental characterization of the PPWG performance in the THz regime in terms of dispersion and field enhancement at the interaction region. We have built several prototypes of the structure out of oxygen-free copper, comprising four identical quadrants that are stacked together as seen in Fig. 3. To be able to characterize the fields at the minimum gap (i.e., the interaction region) we have designed a pocket for an EO crystal to be inserted. Here we have used a 5 mm × 5 mm, 50 μm thick gallium phosphide, (GaP) crystals. A 110-cut GaP crystal is first oriented in free space to align its optical axis with the THz and EO probe beam and then attached to the structure in the pocket. The THz waveforms and field amplitude at the interaction region inside the tapered PPWG were measured using the EOS technique. The THz experimental setup is shown in Fig. 3. THz pulses were generated from 4 mJ, 100 fs laser pulses obtained from an 800 nm Ti:Sapphire laser using the tilted pulse front method [36]. The THz beam was collimated and focused onto the PPWG structure with a pair of 90-degre off-axis parabolic (OAP) mirrors. The polarization of the THz field was perpendicular to the PPWG (along the z-direction in Fig 1). The final 3 inch (76.2 mm) focal length OAP was positioned about 80 mm away from the center of the structure, close to the focal point of the THz beam. To probe the THz field inside structure, a small portion of the 800 nm laser beam was split off and variably delayed in time. The probe beam was spatially and temporally overlapped with the THz beam on the EO crystal in the structure through a hole in the OAP mirror. The polarization state of the probe beam after the EO crystal was analyzed with a quarter wave plate, Wollaston prism and two photo diodes [37].

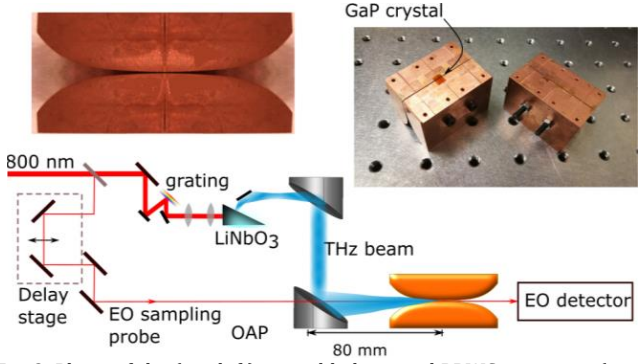

Fig. 3. Photo of the (top left) assembled tapered PPWG structure, (top right) with the EO crystal attached, and (bottom) schematic of the EOS experimental setup for characterizing its THz performance.

In Fig. 4 we show the measured EO waveform compared to the one in free space (for a THz pulse energy of approximately 0.9 μJ) for a PPWG with $g$ = 300 μm. First, we observe that the single cycle characteristics of the THz pulse are preserved, which is an inherent feature of the PPWG. To estimate the local electric field $E_{\text{THz}}$ we use the EO formulas [38]

$$\sin^{-1}(\Delta I / I) = 2\pi n_0^3 r_{41} t_{\text{GaP}} E_{\text{THz}} L / \lambda_0 \qquad (1)$$

where $\Delta I / I$ is the EO modulation index measured by the photodiodes, $r_{41}$ is the electro-optic tensor component of GaP $r_{41}$ = 0.88 pm/V, $L$ is the GaP thickness of 50 μm, $t_{\text{GaP}}$ is the transmission Fresnel coefficient of 0.46 at THz frequencies and $n_0$ = 3.2 is the refractive index at $\lambda_0$ = 800 nm. The measured peak electric field in the PPWG with $g$ = 300 μm is about 300 kV/cm, resulting about 1.55 field enhancement compared to free space. Fig. 4(b) shows the spectrum of both EO traces, each normalized to its own peak, taken using the Fourier transform with a 2 ps window around the peak value of the field. This is done to avoid the artifacts in the measured traces due to THz echoes form the GaP crystal as well as water absorption lines from the ambient air. The spectrum for both the PPWG and free space is centered at 0.68 THz and shows minimal dispersion for the PPWG. We also performed full-wave simulations of the PPWG and the results show also single cycle behavior in the PPWG field, yet a 20% increase in the peak field was calculated compared to the measured data as seen from Fig. 4(c). The peak field values may be reduced in our EOS measurements due to various errors including small uncertainties in EO crystal orientation, distortion of the field due to the EO crystal pocket and the probe beam interaction with the aperture of the copper PPWG especially for small gaps on the order of the EO beam size. Additionally, the THz transverse beam spot asymmetric profile may play a role in the enhancement factor which is not investigated here for brevity.

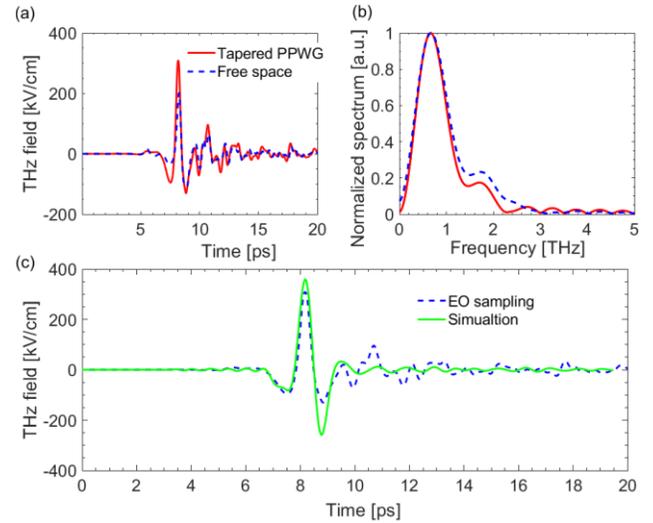

Fig. 4. (a) EO sampling traces of the electric field measured in the PPWG as well as in free space for PPWG with $g$ = 300 μm. (b) The spectrum of the pulses in (a). (c) Comparison between the measured THz electric field in the PPWG compressor and the corresponding simulation for the same incoming pulse energy.

Note that, although field enhancement is an important quality of the structure, the chirp of the electron bunch caused by the structure is the critical design parameter. Therefore, for a given THz pulse energy, we aim to maximize the energy chirp of the beam bunch through the integrated longitudinal electric field observed by the electrons including the transit time in the compressor structure. This can be achieved by tailoring the THz beam spot at the gap using focusing elements or modifying the design as we will discuss in Sec. 4. Several other measurements were also made to characterize the PPWG for different gaps, $g$ = 180 and 450 μm as seen in Fig. 5. Some dispersion is observed for the smallest gap. Here we show both the field enhancement of the PPWG for different gap sizes as well as different distance relative to the OAP focus in Fig. 5. We illustrate the sensitivity of the structure position to relative offsets in $\Delta x$, for the field enhancement maximized at $\Delta x$ = 0.

## 4. Beam Dynamics and a Shorted PPWG

To study the performance merits of the PPWG compressor, i.e., the characteristics of electron beam dynamical interaction with the THz fields, we have developed a time-domain particle tracking method to calculate the electron trajectories. First, we calculate the 3D field maps in the PPWG beam tunnel in the frequency domain within the range 0.05 THz to 1.5 THz based on the simulations outlined in Sec. 2. We transform these fields to time domain using the inverse Fourier transform, and then convolve the results with the measured time domain waveform at the interaction region. We assume an electron bunch modeled as a finite number of macroparticles $N$ defined as point charges and distributed randomly within the bunch spatiotemporal extent. Each macroparticle has its own space-time position and relativistic momentum that are tracked and updated in the given 3D field maps using the Lorentz evolution equations,

$$\frac{d\mathbf{p}_j(t,\mathbf{r}_j(t))}{dt} = q\left[\mathbf{E}_j(t,\mathbf{r}_j(t)) + \mathbf{v}_j(t) \times \mathbf{B}_j(t,\mathbf{r}_j(t))\right] \quad (2)$$

where $j$ = 1,2,...,$N$ is the particle index whose charge is $q$, $\mathbf{p}_j$ is the particles relativistic momentum, while $\mathbf{v}_j$, $\mathbf{E}_j$ and $\mathbf{B}_j$ are the particles velocity, electric and magnetic flux density seen by the particles, respectively (bold represents vector quantities). Here we ignore space-charge forces (inter-charge repulsion) and wakefield effects for simplicity, and we only track particles along the $z$-direction in the beam tunnel considering the transverse size of the beam along the $x$-direction (the PPWG is invariant along the $y$-direction). It is important to emphasize that in the tapered PPWG, the THz propagates perpendicular to the e-beam direction. This causes a transverse deflection (along $x$) from the magnetic field component in the PPWG along $y$. In fact, deflection is a useful property that could be utilized for timing characterization of the bunch (similar to [25]). For deriving the compressor performance, we assume a 5 uJ of available THz energy and an initial 85 fs RMS, 2.5 MeV electron bunch with a 25 μm transverse spot size. These parameters represent typical operating points for use at the UED beamline at SLAC [2]. In Fig. 6(c)-(d) we show the electric and magnetic field at $x = y = 0$ as a function of time and in Fig. 6(e) we show the energy modulation of a single electron passing through the PPWG compressor achieving about 1.3 keV/(100 fs) of energy chirp. This amount of chirp would be sufficient for achieving compression for a beam with a very small transverse size. The peak deflection angle in $x$ shown in Fig. 6(f) is about 2 mrad. However, as the THz waves in the PPWG propagates in the $x$-direction, the time-of-arrival of electrons with respect to the THz pulse would vary significantly along $x$, causing transverse dispersion of the electron beam. This eventually leads to less average chirp across the bunch and consequently reduced compression efficiency. The calculated compression as a function of the THz-to-beam delay reaches a minimum of 45 fs with a compression factor of about 2, shown in Fig. 6(g).

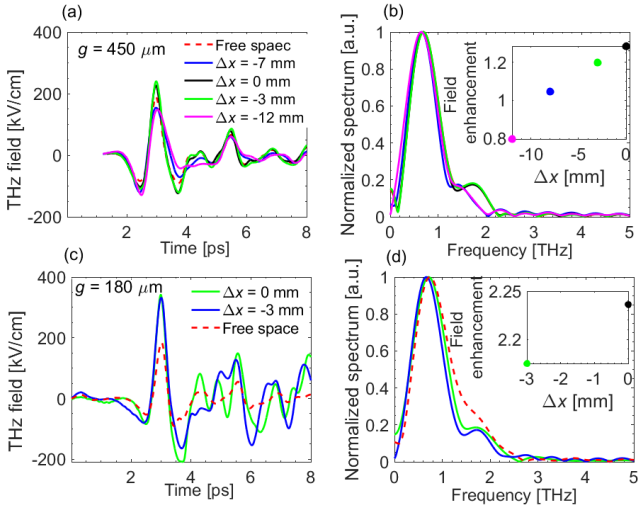

Fig. 5. (a) EO sampling traces measured in the PPWG as well as in free space for $g$ = 450 μm, for different values of $\Delta x$. (b) The spectrum of the pulses in (a). (c) EO sampling traces in the PPWG for $g$ = 180 μm. (b) The spectrum of the pulses in (c). The inset in (b) and (d) shows the field enhancement inside the PPWG as a function of the focal point of the incoming THz pulse.

We propose a new technique to alleviate the transverse effects using a PPWG that is electrically shorted at the location of the beam tunnel at $x = r_b$, see Fig. 6(b). This shorted PPWG produces a standing wave in the interaction region and delivers a uniform electric field across the gap along $x$, compared to the unshorted PPWG. In addition, the magnetic field would also be reduced thanks to the phase acquired by the reflected magnetic field. Furthermore, the peak-to-peak field increases as the electric field constructively add up from the reflection. This is observed in Fig. 6(a)-(d) when compared the two PPWG topologies with the same gap, input THz energy and beam parameters. Accordingly, the shorted PPWG provides a 50% increase in the energy modulation and improves the bunch compression twofold. The energy chirp produced is 2.1 keV/(100 fs) and the minimum bunch length achieved with the shorted structure is about 18 fs with a compression factor of 4.5 that is calculated in 1.2 m downstream. Further compression may be achieved by increasing the THz pulse energy and that would also correspond to compression in a shorter drift distance [31]. Moreover, employing an asymmetrical focusing element such as a lens (or mirror) could provide a tight THz beam waist in both transverse directions at the same $x$-$z$ plane of the PPWG thus enhancing the THz field, however absorption loss of lens must be considered. Finally, we point out that tailoring the spectral phase content of the THz pulse (for instance [39]) would enable better uniformity of the electric field and potentially eliminate the transverse deflection caused by magnetic fields.

### 5. Conclusion

We have analyzed the electrodynamic performance of a new THz-driven compressor design for MeV electron beam bunches using a dispersion free PPWG. We have shown both modelling, and measurement of the THz performance of the structure using EO sampling. The shorted PPWG provides more than a 50% increase in the energy modulation and a twofold enhancement in compression compared to the unshorted tapered PPWG, with a calculated energy chirp of more than 2 keV/100 fs for 5 μJ input THz energy. The THz compressor will serve as a testbed for new time-resolved UED dynamics previously inaccessible with longer electron bunches and could be directly applied to pump-probe investigations of phase transitions and phonon dynamics.

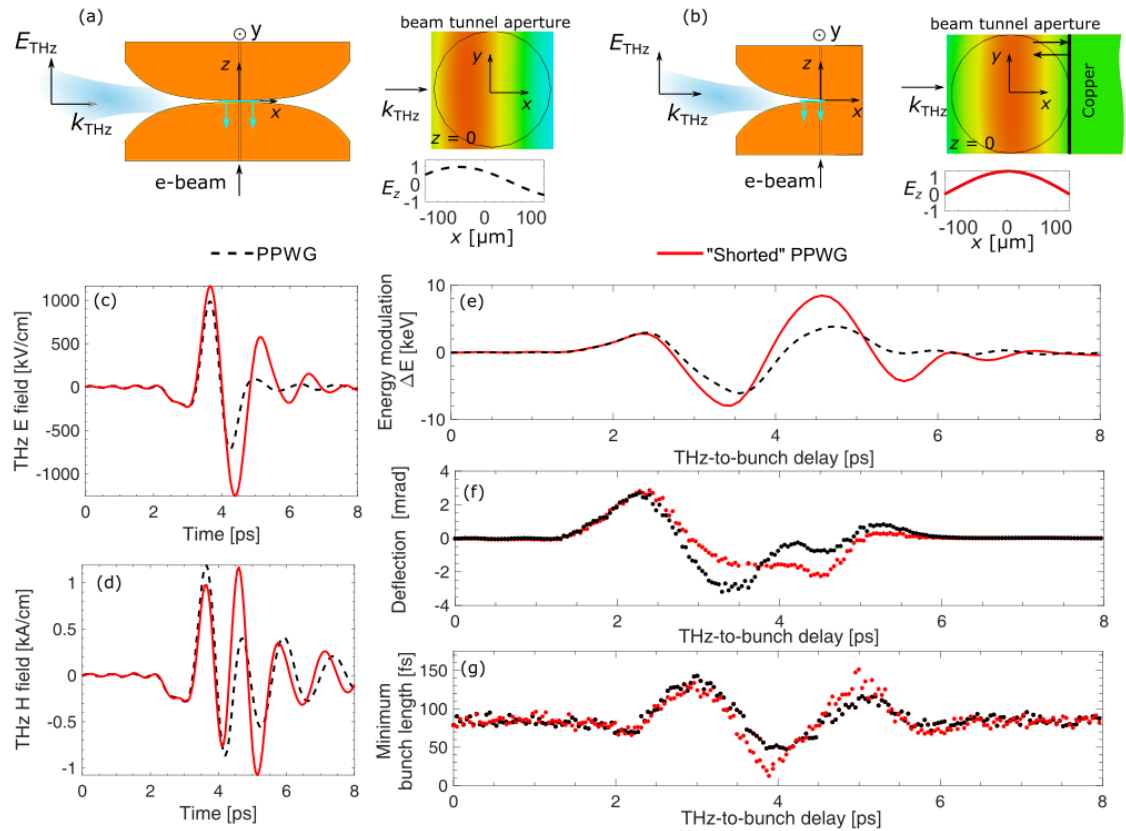

Fig. 6. Simulation results of the performance of the (a) PPWG electron beam compressor compared to (b) the shorted version with the same beam tunnel radius, gap size, and same incoming THz pulse energy of 5 uJ. (c) Electric and (d) magnetic fields in the PPWGs interaction region at $z = 0$ mm and $x = 0$ mm at 0.65 THz. (e) Simulated energy modulation of a 2.5 electron and (f) deflection induced by the PPWG compressor varying as a function of the THz delay for a bunch of 85 fs initial length. (g) Minimum bunch length obtained from the PPWG and the shorted PPWG compressors, in 1.2 m and 0.8 m drift distance respectively.

**Funding**

This work is supported by US Department of Energy (DOE) Contract No. DE-AC02-76SF00515. UED is supported in part by DOE BES Scientific User Facilities Division and SLAC UED/UEM program development: DE-AC02-05CH11231. Use of the Linac Coherent Light Source, SLAC National Accelerator Laboratory, is supported by the US DoE, Office of Science, Office of Basic Energy Sciences under contract numbers DE-AC02-76SF00515. M.K. and M.C.H. are supported by the US Department of Energy, Office of Science, Office of Basic Energy Sciences, under award no. 2015-SLAC-100238-Funding.

**Acknowledgment**

The authors would like to thank Mario Cardoso, Ann Sy, and, Andrew Haase for structure fabrication, Emma Snively and Xiaozhe Shen for useful discussions.